\documentclass[runningheads]{llncs}
\usepackage{graphicx}

\usepackage{tikz}
\usepackage{comment}
\usepackage{amsmath,amssymb} 
\usepackage{color}
\usepackage{bbm}

\usepackage[accsupp]{axessibility}  

\begin{document}
\pagestyle{headings}
\mainmatter
\def\ECCVSubNumber{100}  

\title{FDVTS's Solution for 2nd COV19D Competition on COVID-19 Detection and Severity Analysis} 


\titlerunning{Abbreviated paper title}
%
\author{Junlin Hou \and Jilan Xu \and Rui Feng \and Yuejie Zhang}
\authorrunning{F. Author et al.}
%
\institute{School of Computer Science, Shanghai Key Lab of Intelligent Information Processing, Fudan University\\
\email{\{jlhou18,jilanxu18,ruifeng,yjzhang\}@fudan.edu.cn}\\
}
\maketitle

\begin{abstract}
This paper presents our solution for the 2nd COVID-19 Competition, occurring in the framework of the AIMIA Workshop in the European Conference
on Computer Vision (ECCV 2022). In our approach, we employ an effective 3D Contrastive Mixup Classification network for COVID-19 diagnosis on chest CT images, which is composed of contrastive representation learning and mixup classification. For the COVID-19 detection challenge, our approach reaches 0.933 macro F1 score on 484 validation CT scans, which significantly outperforms the baseline method by 16.3\%. In the COVID-19 severity detection challenge, our approach achieves 0.770 macro F1 score on 61 validation samples, which also surpasses the baseline by 14\%.

\end{abstract}

\section{Introduction} 
\label{section:sec1}
The Coronavirus Disease 2019 SARS-CoV-2 (COVID-19) is a highly infectious disease, which emerged in December, 2019 \cite{WHO}. Early detection based on chest CT scans is important to the timely treatment of patients and the slowdown of viral transmission.
However, a volumn of CT scans contains hundreds of slices, which requires a heavy workload on radiologists. 
Recently, deep learning approaches have achieved excellent performance in fighting against COVID-19. They have been widely applied to many aspects, including the lung and infection region segmentation \cite{weakly,li2020artificial,Chen2020.02.25.20021568,tsinghua2020fourweek} as well as the clinical diagnosis and assessment \cite{wang2020automatically,wang2020a,song2020deep,HOU2021108005}. 

In this paper, we present deep learning based solution for the 2nd COV19D Competition of the Workshop ``AI-enabled Medical Image Analysis – Digital Pathology \& Radiology/COVID19 (AIMIA)'', which occurs in conjunction with the European Conference on Computer Vision (ECCV) 2022.
The competition includes two challenges, namely COVID-19 Detection Challenge and COVID-19 Severity Detection Challenge. COVID-19 detection aims to identify COVID from non-COVID cases. Each CT scan is manually annotated with respect to COVID-19 and non-COVID-19 categories. The severity of COVID-19 can be further divided into four stages, including Mild, Moderate, Severe, and Critical. Each severity category is determined by the presence of ground glass opacities and the pulmonary
parenchymal involvement.

A natural way to diagnose COVID-19 based on 3D CT images is to use 3D networks. To address both challenges, we employ the advanced 3D contrastive mixup classification network (CMC-COV19D) in our previous work \cite{hou2021cmc}, which won the first price in the ICCV 2021 COVID-19 Diagnosis Competition of AI-enabled Medical Image Analysis Workshop \cite{kollias2021mia}. Our CMC-COV19D framework introduces contrastive representation learning to discover more discriminative representations of COVID-19 cases. Besides, we use a joint training loss that combines the classification loss, mixup loss, and contrastive loss. We further employ an inflated 3D ImageNet pre-trained ResNest50 \cite{zhang2020resnest} as a strong feature extractor to boost more accurate COVID-19 diagnostic performance. Experimental results on both challenges show that our approach significantly surpasses the baseline model provided by organizers. 


\section{Dataset} \label{section:dataset}

COV19-CT-DB \cite{kollias2022ai} includes 3D chest CT scans annotated for existence of COVID-19. It consists of 1,650 COVID and 6,100 non-COVID chest CT scan series, which correspond to a high number of patients (more than 1150) and subjects (more than 2600). In total, 724,273 slices correspond to the CT scans of the COVID-19 category and 1,775,727 slices correspond to the non COVID-19 category. Each of the 3D scans includes different number of slices, ranging from 50 to 700.

The database has been split in training, validation and testing sets. The training set contains 1992 3D CT scans. The validation set consists of 494 3D CT scans. A further split of the COVID-19 cases has been implemented, based on the severity of COVID-19, in the range from 1 to 4. In particular, parts of the COV19-CT-DB COVID-19 training and validation datasets have been accordingly split for severity classification in training, validation and testing sets. The training set contains, in total, 258 3D CT scans. The validation set consists of 61 3D CT scans.


\section{Methodology} \label{section:method}

\begin{figure}
\centering
\includegraphics[width=\textwidth]{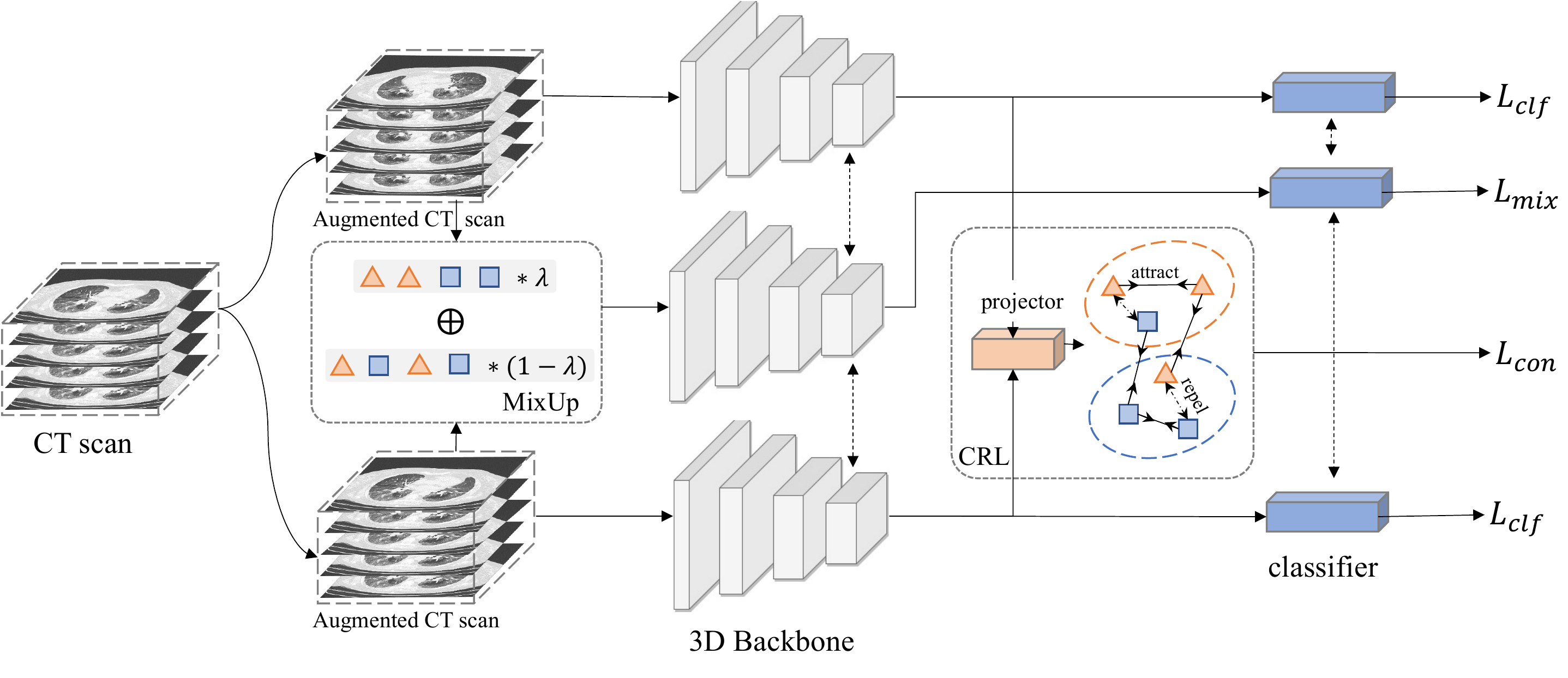}
\caption{Overview of our CMC-COV19D Network.}
\label{fig:cmc}
\end{figure}

As shown in Fig. \ref{fig:cmc}, Our CMC-COV19D network is composed of contrastive representation learning (CRL) and mixup classification.

\subsubsection{Contrastive Representation Learning.}
Our CMC-COV19D network employs the contrastive representation learning (CRL) as an auxiliary task to learn more discriminative representations of COVID-19. The CRL is comprised of the following components. 1) A stochastic data augmentation module $A(\cdot)$, which transforms an input CT $x$ into a randomly augmented sample $\tilde{x}$. We generate two augmented volumes from each input CT. 2) A base encoder $E(\cdot)$, which maps the augmented CT sample $\tilde{x}$ to a representation vector $r=E(\tilde{x})\in \mathbb{R}^{d_e}$. 3) A projection network $P(\cdot)$, which is used to map the representation vector $r$ to a relative low-dimension vector $z=P(r)\in \mathbb{R}^{d_p}$. 4) A classifier $C(\cdot)$, which classifies the vector $r\in \mathbb{R}^{d_e}$ to the final prediction. 

Given a minibatch of $N$ CT images and their labels $\{(x_k,y_k)\}_{k=1,\dots,N}$, we can generate a minibatch of $2N$ samples after data augmentations. 
Inspired by the supervised contrastive loss \cite{2020Supervisedcon}, we define the positives as any augmented CT samples from the same category, whereas the CT samples from different classes are considered as negative pairs. Let $i\in \{1,\dots,2N\}$ be the index of an arbitrary augmented sample, the contrastive loss function is defined as:
\begin{equation}
    \mathcal{L}_{con}^i=\frac{-1}{2N_{\tilde{y}_i}-1}\sum_{j=1}^{2N}\mathbbm{1}_{i\ne j}\cdot\mathbbm{1}_{\tilde{y}_i=\tilde{y}_j}\cdot\log\frac{\exp(z_i^T\cdot z_j/\tau)}{\sum_{k=1}^{2N}\mathbbm{1}_{i\ne k}\cdot\exp(z_i^T\cdot z_k/\tau)},
\label{infonce}
\end{equation}
where $\mathbbm{1}\in\{0,1\}$ is an indicator function, and $\tau >0$ denotes a scalar temperature hyper-parameter. $N_{\tilde{y}_i}$ is the total number of samples in a minibatch that have the same label $\tilde{y}_i$. 

\subsubsection{Mixup classification}

We adopt the mixup \cite{zhang2017mixup} strategy during training to further boost the generalization ability of the model. For each augmented CT sample $\tilde{x}_i$, we generate the mixup sample and it label as:
\begin{equation}
    \tilde{x}^{mix}_i = \lambda\tilde{x}_i + (1-\lambda)\tilde{x}_{p}, 
    ~\tilde{y}^{mix}_i = \lambda\tilde{y}_i + (1-\lambda)\tilde{y}_{p}, 
\end{equation}
where $p$ is randomly selected indice. The mixup loss is defined as the cross entropy loss of mixup samples:
\begin{equation}
    \mathcal{L}^i_{mix} = \mathrm{CrossEntropy}(\tilde{x}^{mix}_i,\tilde{y}^{mix}_i).
\end{equation}
Different from the original design \cite{zhang2017mixup} where they replaced the classification loss with the mixup loss, we merge the mixup loss with the classification loss to enhance the classification ability on both raw samples and mixup samples. The classification loss $\mathcal{L}_{ce}$ on raw samples is defined as:
\begin{equation}
    \mathcal{L}_{clf}^i=-\tilde{y}_i^T\log\hat{y}_i,
\end{equation}
where $\tilde{y}_i$ denotes the one-hot vector of ground truth label, and $\hat{y}_i$ is predicted probability of the sample $x_i$ $(i=1,\dots,2N)$.

Finally, we merge the CRL loss, mixup loss, and classification loss into a combined objective function:
\begin{equation}
     \mathcal{L}=\frac{1}{2N}\sum_{i=1}^{2N}(w_1\mathcal{L}^i_{con}+w_2\mathcal{L}^i_{mix}+w_3\mathcal{L}^i_{clf}),
\end{equation}
where $w_1, w_2, w_3$ denote the balance weights.


\section{Experimental Results} \label{section:experiments}

\subsection{Implementation Details}




For the COVID-19 detection, each CT volume is resized from $(N,512,512)$ to $(128,256,256)$, where $N$ denotes the number of slices. Data augmentation includes RandomResizedCrop, random crop on the z-axis to 64, random contrast changes. Other augmentations such as flip and rotation are also tried, but no significant improvement is yielded. 

For the COVID-19 severity detection, we resample each CT volume into $(64,256,256)$. The same augmentations are applied, except for random crop on the z-axis.

We employ inflated 3D ResNest50 and Uniformer-S as the backbones in our experiments. The networks are trained for 100 epochs. 
We optimize the networks using the Adam algorithm with a weight decay of $10^{-5}$. The initial learning rate is set to $10^{-5}$ and then divided by 10 at $30\%$ and $80\%$ of the total number of training epochs. Our methods are implemented in PyTorch and run on four NVIDIA Tesla V100 GPUs. 

We adopt the Macro F1 score as the evaluation metric. The score is defined as the unweighted average of the class-wise/label-wise F1 Scores.

\subsection{Results on the COVID-19 detection challenge}

\setlength{\tabcolsep}{4pt}
\begin{table}[t]
\begin{center}
\caption{The results on the validation set of COVID-19 detection challenge.* indicates adaptive mixup and cutmix strategy.}
\label{table:covid detection}
\begin{tabular}{llcccc}
\hline\noalign{\smallskip}
ID & Backbone & Pretrain & CRL & Mixup & macro F1 score\\
\noalign{\smallskip}
\hline
\noalign{\smallskip}
1& CNN-RNN \cite{kollias2022ai} & -& - & - & 0.770\\
2& ResNest50 & ImageNet &  &  & 0.897\\
3& ResNest50 & ImageNet & $\checkmark$ & $\checkmark$ & 0.924\\
4& Uniformer-S & ImageNet &  &  & 0.915\\
5& Uniformer-S & ImageNet & $\checkmark$ & $\checkmark$  & 0.920 \\
6& Uniformer-S & k400 & $\checkmark$ & $\checkmark$  & 0.925 \\
7& Uniformer-S & k400\_16x8 & $\checkmark$ & $\checkmark$  & 0.927 \\
8& Uniformer-S* & k400\_16x8 & $\checkmark$ & $\checkmark$  & 0.933 \\
\hline
\end{tabular}
\end{center}
\end{table}
\setlength{\tabcolsep}{1.4pt}

Table \ref{table:covid detection} shows the results of the baseline model and our methods on the validation set of COVID-19 detection challenge. The baseline CNN-RNN approach follows the work~\cite{kollias2020deep,kollias2020transparent,kollias2018deep} on developing deep neural architectures for predicting COVID-19. It achieves 0.77 macro F1 score. Our 3D CMC-COV19D models with different backbones obtain significant improvements compared with the baseline. The effectiveness of CRL and Mixup modules are also verified.


\subsection{Results on the COVID-19 severity detection challenge}

\setlength{\tabcolsep}{4pt}
\begin{table}[t]
\begin{center}
\caption{The results on the validation set of COVID-19 severity detection challenge.}
\label{table: severity}
\begin{tabular}{llcccc}
\hline\noalign{\smallskip}
ID& Methods & CRL & Mixup & Macro F1 score\\
\noalign{\smallskip}
\hline
\noalign{\smallskip}
1& CNN-RNN \cite{kollias2022ai}& -& - & 0.630\\
2& ResNest50 &  &  &  0.655\\
3& ResNest50 & & $\checkmark$ &  0.673\\
4& ResNest50 & $\checkmark$ & $\checkmark$ & 0.719\\
5& ResNest50+Lesion & $\checkmark$ & $\checkmark$ & 0.770\\
\hline
\end{tabular}
\end{center}
\end{table}
\setlength{\tabcolsep}{1.4pt}

Table \ref{table: severity} shows the results of the baseline model and our methods on the validation set of COVID-19 severity detection challenge. The baseline CNN-RNN approach achieves 0.63 macro F1 score. As can be seen from the 2nd to 4th rows of Table \ref{table: severity}, our 3D CMC-COV19D models also achieve better performance.

\section{Conclusions} \label{section:conclusion}
In this paper, we present our solution for the 2nd COVID-19 Competition on two challenges: COVID-19 detection and COVID-19 severity detection. 
Our network is composed of contrastive representation learning and mixup classification for more accurate COVID-19 diagnosis. We achieve 0.933 and 0.770 macro F1 score on the the validation set of COVID-19 detection and severity detection, respectively.

%
%
\bibliographystyle{splncs04}
\bibliography{egbib}
\end{document}